\documentclass[global,twocolumn]{svjour}
\usepackage{graphics}
\journalname{Applied Physics B}
\begin{document}
\title{Narrow linewidth single laser source system for onboard atom interferometry}
\author{Fabien Theron, Olivier Carraz\thanks{\emph{Present address:} European Space Agency - ESTEC Future Missions Division (EOP-SF) P.O. Box 299, 2200 AG Noordwijk, The
Netherlands.}, Geoffrey Renon, Nassim Zahzam, Yannick Bidel, Malo Cadoret\thanks{\emph{Present address:} Laboratoire Commun de M´etrologie CNAM, 61 rue du Landy, 93210 La Plaine Saint Denis, France.}, Alexandre Bresson
}                     % Do not remove
\institute{ONERA - The French Aerospace Lab, BP 80100, 91123 Palaiseau Cedex, France,
E-mail: fabien.theron@onera.fr, Fax: +33 1 8038 6182}
%
%\offprints {hnamei}
%\mail {fabien.theron@onera.fr}
\date{Received: date / Revised version: date}
\maketitle
\begin{abstract}
A compact and robust laser system for atom interferometry based on a frequency-doubled
telecom laser is presented. Thanks to an original stabilization architecture on a saturated absorption setup, we obtain a
frequency-agile laser system allowing fast tuning of the laser frequency over 1 GHz in few ms using 
a single laser source. The different laser frequencies used for atom interferometry are generated by changing
dynamically the frequency of the laser and by creating sidebands using a phase modulator. A laser
system for Rubidium 87 atom interferometry using only one laser source based on a frequency doubled telecom fiber
bench is then built. We take advantage of the maturity of fiber telecom technology to reduce the number of free-space optical components
(which are intrinsically less stable) and to make the setup compact and much less sensitive to vibrations and
thermal fluctuations. This source provides spectral linewidth below 2.5 kHz, which is required for precision atom
interferometry, and particularly for a high performance atomic inertial sensor.
\end{abstract}

\section{Introduction}
\label{intro}
Atom interferometers have demonstrated excellent performances for precision acceleration and rotation
measurements \cite{borde,precision}. Many applications of these sensors, as tests of fundamental physics in space \cite{fundamental} or
gravimetry \cite{girafe}, need the setup to be compact, transportable and robust in order to operate in relevant
environments (satellite, planes, boats). Intensive research has been carried out over the last few years to develop transportable laser systems
\cite{carraz,tino,syrte,laserice} and, in particular, laser systems as compact and immune to perturbations as possible. 

Atom interferometers usually operate with alkali atoms by driving transitions in the near-IR spectrum (852 nm for
Cs, 780 nm for Rb, and 767 nm for K). A light pulse atom interferometer sequence consists usually of three stages. First, a gas of
atoms is cooled, trapped and selected in a non sensitive magnetic state. Second, these cold atoms are illuminated by a sequence of
three light pulses driving stimulated Raman transitions performing a Mach-Zehnder type atom interferometer \cite{raman}. Finally,
the phase shift of the interferometer is deduced from fluorescence measurements. In these experiments, we need to use several stable laser
frequencies relative to the atomic transitions of the alkali species considered: one cooling and trapping frequency, one repumping frequency, two Raman
frequencies and one detection frequency. It is noted that the spectral linewidth of the laser needs to be smaller than the linewidth of the atomic transition
for the cooling stage (6 MHz for Rb). Additionally, in order to realize stimulated Raman transitions, an even narrower linewidth laser is required because the frequency
noise of the laser induces a noise on the atom interferometer measurement \cite{sensitivity}. This aspect is very important for gravity
gradiometers for which the sensitivity is not limited by vibrations.

\section{State of the art}
Different technologies of laser sources are available for addressing alkali atoms. For example, laser sources emitting directly at the same wavelength as the
atomic transition can be used: Distributed FeedBack lasers (DFB), Distributed Bragg Reflector lasers (DBR) and Extended-Cavity Diode Lasers
(ECDL) \cite{diode}. However, the disadvantage of these technologies is that large efforts are required to obtain robust systems, immune to 
mechanical misalignments caused by vibrations, for onboard applications. Another appropriate solution for Rb and K,
is to use frequency-doubled telecom lasers operating around 1.5 $\mu$m \cite{lienhart,potassium}. This technique is based on the
maturity of the fiber components in the telecom C-band to reduce the amount of free-space optics and to make the setup more compact
and less sensitive to misalignments. Moreover, many types of narrow linewidth laser sources are commercially available such as DFB laser,
DFB with whispering-gallery-mode resonator \cite{WGM}, integrated ECDL diodes \cite{RIO} and Erbium Fiber DFB Laser (EFL).
In this article, we will present results obtained with an EFL source which has already been used for atom interferometry \cite{carraz}.

Different architectures are possible to obtain all the laser frequencies needed for an atom interferometer experiment.
The most common one uses at least two lasers: a master laser and a slave laser \cite{carraz,tino,syrte}. The master laser
has a fixed frequency and is locked on an atomic transition. The slave laser is locked relatively to the master laser thanks to
a beat note. By changing the set point of the beat note lock, it is possible to change dynamically the frequency of the
slave laser and to address all the functions needed for atom interferometry. However, for onboard applications where a 
compact and robust laser system is needed, the use of two laser sources is not optimal. Indeed, by using only
one laser source, the size of the laser system is limited, the risk of failure of the system due to laser source breakdown is
reduced and the electrical consumption is lower.

\section{Description of the laser system}
\subsection{Laser system}
\begin{figure}[h!]
\centerline{\scalebox{0.30}{\includegraphics{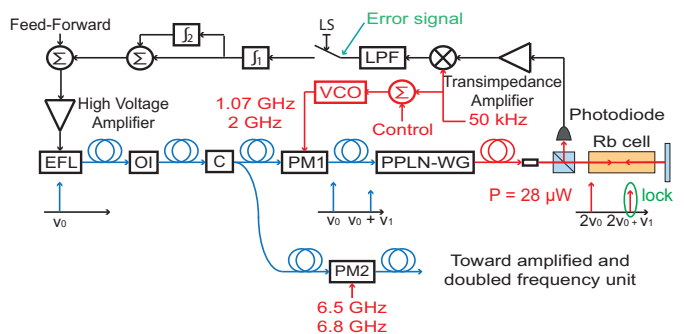}}}
\caption{Diagram of the laser system and the electronic lock : EFL, Erbium Fiber DFB Laser; OI, Optical Isolator; PM, Phase Modulator; C, fiber
Coupler; PPLN-WG, Periodically Poled Lithium Niobate crystal - Wave Guide; VCO, Voltage Controlled Oscillator; LPF, Low Pass Filter; LS, Lock
Switch (open during the frequency step); $\nu_{0}$, optical EFL frequency; $\nu_{1}$, modulation frequency of the PM1.}
\label{laser}
\end{figure}

\begin{figure}[h!]
\centerline{\scalebox{0.40}{\includegraphics{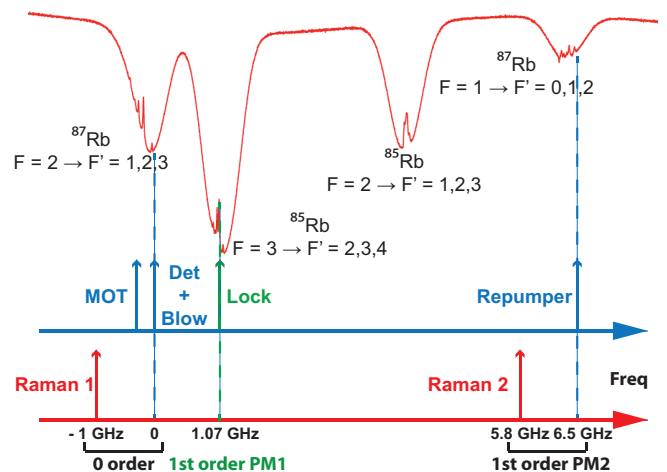}}}
\caption{Saturated absorption peaks of the D$_2$ Rubidium transition and laser frequencies generated for atom
interferometry (laser frequencies for the cooling and detection stages in blue, and for the interferometric stage in red).}
\label{absorption}
\end{figure}

In this article, we present a laser system for Rubidium 87 atom interferometry using only one laser source based on a
frequency-doubled telecom fiber bench. A laser system, tunable in few ms, within a frequency range of typically
1 GHz, can generate cooling, detection and the first Raman frequencies. The repumping frequency and the second Raman frequency
can be obtained by creating sidebands on the laser source. In our system, the laser source is a narrow linewidth EFL (IDIL fiber laser, output power: 20 mW, linewidth $<$ 2 kHz), which can be frequency tuned (20 MHz/V) thanks to a piezoelectric actuator (PZT) (Fig.~\ref{laser}). In order to frequency stabilize the laser, part of the laser output goes through a phase modulator (PM1, Photline, RF level: 23 dBm) which generates side bands, then goes through a PPLN waveguide crystal (NTT Electronics, conversion efficiency: 225 \%/W) which performs second harmonic frequency conversion, and finally goes in a Rubidium saturated absorption setup \cite{absorption} where the 1$^{st}$ order sideband of the laser spectrum is locked
to the cross over $F=3\rightarrow{}F'=3 \ c.o.\ 4$ of the $^{85}$Rb-D$_2$ line (Fig.~\ref{absorption}). With a modulation at $\nu_{1} =$ 1070 MHz,
the laser carrier (0$^{th}$ order sideband) is at resonance with the detection transition $F=2\rightarrow{}F'=3$ of the $^{87}$Rb-D$_2$ line. By changing
the frequency modulation $\nu_{1}$ on PM1, the 1$^{st}$ order sideband of the laser remains locked on the atomic transition whereas the frequency of the carrier is varied. In that case, the new carrier is detuned relative to the detection transition by 1070 MHz - $\nu_{1}$. In summary, the 1$^{st}$ order sideband is locked on the deepest saturated absorption peak of the Rubidium 85, while the carrier is tuned to the Rubidium 87 transitions.
With this technique, a frequency tuning range of at least 1 GHz at 780 nm can be obtained, which therefore addresses all the functions needed for atom interferometry.

\subsection{Frequency stabilization}
\begin{figure}[h!]
\centerline{\scalebox{0.30}{\includegraphics{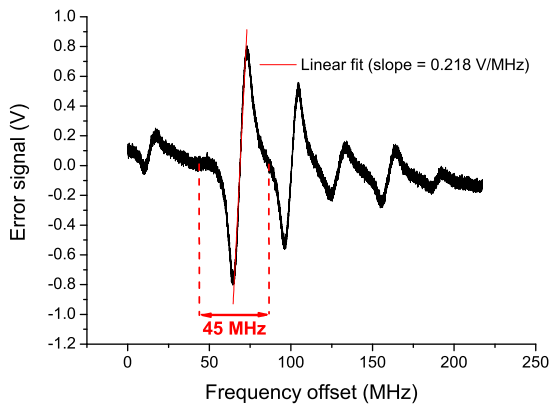}}}
\caption{Error signal as a function of the frequency scan of the laser (@ 780 nm).}
\label{error}
\end{figure}

Frequency stabilization is achieved by modulating at 50 kHz the frequency $\nu_{1}$ driving the PM1, and by collecting saturated
absorption signal with a photodiode. The laser beam used for this saturated absorption has a power equal to 28 $\mu$W @780 nm (3.56 mW @1560 nm before frequency doubling) with a beam diameter of 1.8 mm. This signal is then amplified through a transimpedance amplifier, demodulated at 50 kHz and low-pass filtered, hence providing at this point the dispersive error signal of the lock system proportional to the frequency difference between the 1$^{st}$ order sideband of the laser spectrum and the atomic frequency transition.
Finally, it is integrated and amplified by a high-voltage amplifier, and sent to the PZT of the EFL with a feedback bandwidth of 3 kHz
(Fig.~\ref{laser}). The amplitude of the 50 kHz modulation is adjusted in order to have
 the steepest slope (0.218 V/MHz).
With this architecture, a peak-to-peak deviation of 7.8 MHz, and a capture capture range of 45 MHz on the error signal is obtain (Fig.~\ref{error}). This finite range is due to the presence of absorption peaks located before and after the peak where we are locked.

\begin{figure}[h!]
\centerline{\scalebox{0.30}{\includegraphics{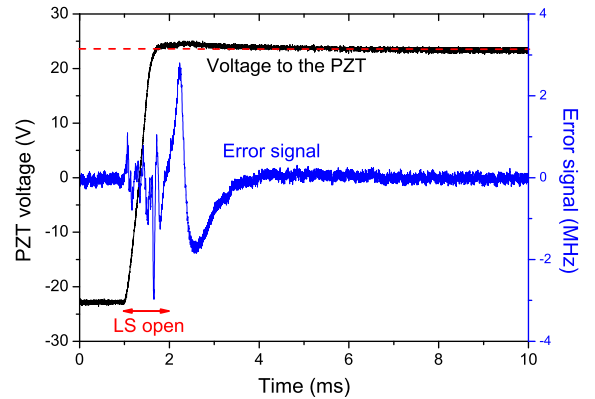}}}
\caption{Behavior of the laser system locked during a frequency step : voltage to the PZT of the EFL (black);
error signal (blue). Voltage step of 46.4 V on the PZT, i.e. frequency step of 965.12 MHz on the laser at 780 nm.}
\label{raman}
\end{figure}

In order to improve the stability and reduce the response time of the lock system during frequency steps, a feedforward (Fig.~\ref{laser}) proportional to the frequency of the VCO is added to the corrected signal driving the PZT. During the maximum frequency step of 1 GHz, the frequency deviation is too large compared to the lock range for the laser to remain locked. Therefore, a lock switch (ADG201A) is upstream from the integrator to open the feedback loop during 1 ms after the frequency step. A second
integrator stage is also added to reduce the stabilization time. As a result, after a frequency step of 1 GHz, the laser frequency is stabilized with an error below 100 kHz 
after 3 ms only (Fig.~\ref{raman}).

With this laser architecture, to obtain the second Raman frequency and the repumper required to realize the atom interferometry experiment
(Fig.~\ref{absorption}), the laser is modulated at a frequency of 6.5 or
6.8 GHz with the PM2 (Photline, RF level from -1 to 23 dBm) (Fig.~\ref{laser}) \cite{parasite}. Finally, the laser is
amplified in an Erbium Doped Fiber Amplifier (IPG Photonics, input power: 3 mW, output power: 5 W) and sent in a frequency-doubling unit. The frequency doubling can be implemented either with a PPLN waveguide \cite{CNES} or with free-space doubling in a bulk PPLN \cite{carraz}.

\section{Frequency noise and influence on the atom interferometer}
\subsection{Estimation of the frequency noise of the laser}
\begin{figure}[h!]
\centerline{\scalebox{0.30}{\includegraphics{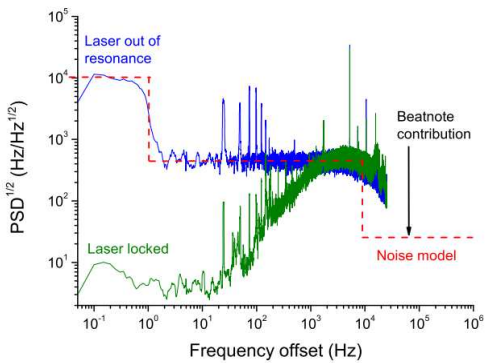}}}
\caption{Square root of the Power Spectral Density (PSD) of the error signal noise : when the laser is locked (green) and when the laser is out of resonance (blue). The noise of the error signal when the laser is out of resonance (blue) determines the low frequency part ($f <$ 10 kHz) of the frequency noise. The red dash line represents the model used to determine the atom interferometry sensitivity.}
\label{model}
\end{figure}

\begin{figure}[h!]
\centerline{\scalebox{0.30}{\includegraphics{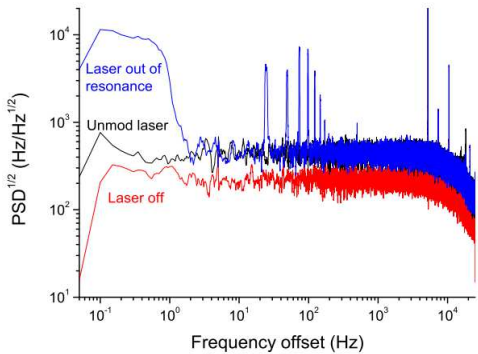}}}
\caption{Square root of the PSD of the error signal noise : when the laser is off (red), when the laser is unmodulated (black) and when the laser is out of resonance (blue).}
\label{bruit_laser}
\end{figure}

In order to determine the frequency noise of the laser, it is necessary to analyse the noise of the error signal (Fig.~\ref{model}), which is
proportional to the laser frequency within a bandwdith of 10 kHz (cut-off frequency of the LPF on Fig.~\ref{laser}). First, this noise is measured when the laser is unlocked
and "out of resonance" from the atomic transition (in blue). In this configuration only the noise of our lock system (i.e. the noise
on the error signal which does not come from the frequency noise of the laser) is measured. When the laser is "locked" (in green), the noise on the error signal
is much lower than the noise of the lock. As a result, the frequency noise of the laser is given by the frequency noise of the lock system up to a frequency of 3 kHz.
We investigate then the origin of the noise of the lock system which determines the frequency noise of the laser. The noise of the lock system can come mainly from electronic noise, intensity noise of the laser at 50 kHz and etalon effects in the saturated absorption setup which lead to a temporal fluctuation of the error signal offset.
The origin of the noise can be determined by analyzing the noise of the error signal for different configurations (Fig.~\ref{bruit_laser}). In the configuration
"laser off" (in red), the only contribution comes from the electronic noise. In the configuration laser on, out of resonance and PM1 off ("unmod laser" in black), both the electronic noise and the intensity noise of the laser are present. In the configuration laser "out of resonance" (in blue), all the noise sources are taken into account. Comparing these
configurations shows that the noise of the lock system comes mainly from the intensity noise of the laser between 1 Hz and 10 kHz, whereas the noise comes from etalon effects below 1 Hz. As a summary, the frequency noise of our laser below 10 kHz comes from the noise of the lock which is converted into frequency noise in the feedback loop.
The noise of the lock is mainly due to intensity noise of the laser and etalon effects in the saturated absorption setup.

\begin{figure}[h!]
\centerline{\scalebox{0.30}{\includegraphics{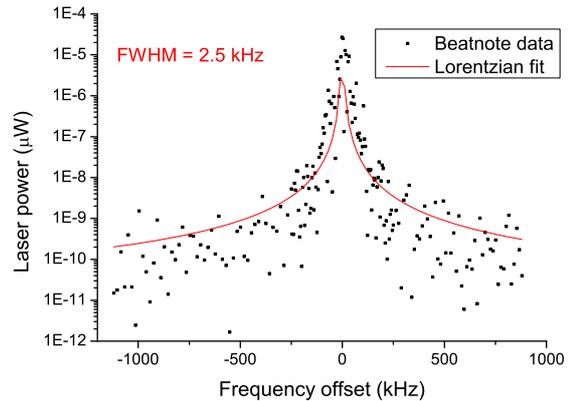}}}
\caption{Beat note between the EFL and an integrated ECDL with a linewidth lower than 10 kHz. The beat note determines the high frequency part ($f >$ 10 kHz) of the frequency noise.}
\label{beatnote}
\end{figure}

In order to estimate the frequency noise of our laser at frequencies higher than 10 kHz, we perform a beat note measurement between our locked EFL and an integrated
ECDL (RIO ORION laser source) which linewidth is below 10 kHz according to the manufacturer.
Because the linewidth of the ECDL is not infinitely narrow, the analysis of the beatnote gives an upper limit of the frequency noise of the EFL. We notice that the wings of the
beatnote fit well with a lorentzian function with a FWHM of 2.5 kHz (Fig.~\ref{beatnote}). As we know that a white frequency noise leads to a lorentzian spectrum with a FWHM equal to $\pi$.S$_{\nu}^{0}$, we suppose that the frequency PSD of our laser at high frequencies is below $\Delta\nu/\pi$. Thus, for the determination of the atom interferometry noise, we will consider that the frequency noise of our laser is equal to (S$_{\nu}^{0})^{1/2} = 28$ Hz/Hz$^{1/2}$ for frequencies above 10 kHz.

\subsection{Estimation of atom interferometry noise induced by the laser frequency noise}
From the two previous measurements,
we can model our laser spectrum by (S$_{\nu}^{0})^{1/2}$ = 10$^{4}$ Hz/Hz$^{1/2}$ for $f <$ 1 Hz, coming from etalon effects, (S$_{\nu}^{0})^{1/2} = 400$ Hz/Hz$^{1/2}$
for 1 Hz $< f <$ 10 kHz, due to intensity noise, and (S$_{\nu}^{0})^{1/2} = 28$ Hz/Hz$^{1/2}$ for $f >$ 10 kHz (red dash line in Fig.~\ref{model}).
Therefore we can estimate the noise on the atom interferometer measurement induced by the frequency noise of the laser.
From the results of \cite{sensitivity} and considering classical experimental parameters for a vertical
atom accelerometer (the distance atom-mirror for gravimetry, or the distance between the two atom clouds for gradiometry: L = 1 m, the duration of a Raman pulse: $\tau_{R}$ = 10  $\mu$s, and the time between two pulses T = 100 ms), a single shot rms noise equal to $\sigma_{a} =  2.6 \times 10^{-9}$ g is obtained, the main contribution coming from low
frequency noise between 1 Hz and 10 kHz. 

\section{Improvements}
For more demanding applications, frequency noise could be decreased by improving the saturated absorption setup (i.e. removing intensity noise of
the laser and etalon effect coming from the fiber \cite{abs_diff}). The electronic noise could also be decreased by using a higher laser power or a more efficient
lock system. With these improvements,  it should be possible
to obtain a frequency noise of 28 Hz/Hz$^{1/2}$ over the whole frequency range of the laser, leading to single shot rms noise equal to $\sigma_{a} = 6.9 \times 10^{-10}$ g for typical atom accelerometers. 
In our laser architecture, the EFL could also be replaced by an integrated ECDL which is more compact and has comparable laser linewidth. Finally, a very compact
system, immune to external disturbances could be built from an all-fibered bench with a fibered amplifier and a wave guided PPLN after the PM2 \cite{CNES}.

\section{Conclusion}
We have developed a tunable narrow linewidth single laser source system for atom interferometry.
This system combines the  reliability of fiber components and the agility allowed by phase modulators. These features can lead to
plug-and-play laser sources for laboratories developing cold atom experiments. These sources could be developed for 
commercial devices, onboard system or space missions. Finally, the use of a single laser source significantly
reduces failure risks and the amount of additional components for redundancy, particularly critical for space projects.\\

\textbf{\scriptsize{ACKNOWLEDGEMENTS}} \, We thank F. Nez, from the Laboratoire Kastler Brossel (LKB), for his help on the project.
We acknowledge funding support from the Direction Scientifique G\'en\'erale of ONERA, the Direction G\'en\'erale de l'Armement (DGA), and the Centre National d'Etudes Spatiales (CNES).

\end{document}